\documentclass[technote,10pt]{IEEEtran}

\usepackage{amsmath,bm,amssymb}
\usepackage{cases}

\usepackage{graphicx}
\usepackage[font=small,skip=0pt]{caption} 
\usepackage{epstopdf}

\usepackage{cite}
\usepackage{hyperref}

\usepackage{amsthm}
\theoremstyle{plain}
\newtheorem{lemma}{Lemma}
\newtheorem{theorem}{Theorem}

\usepackage{setspace}
\usepackage{relsize}

\DeclareMathSizes{10}{9}{6}{4} 

\begin{document}

\title{User-Centric Joint Transmission in Virtual-Cell-Based Ultra-Dense Networks}
	
\author{
	Yingxiao~Zhang, Suzhi~Bi,
	 ~and Ying-Jun Angela Zhang
	\thanks{
		\IEEEcompsocthanksitem
		\copyright2017 IEEE. Personal use of this material is permitted. However, permission to use this material for any other purposes must be obtained from the IEEE by sending a request to pubs-permissions@ieee.org. 
		\IEEEcompsocthanksitem
		This work was supported in part by General Research Funding (Project No. 14209414) from the Research Grants Council of Hong Kong and the National Basic Research Program (973 program Program No. 2013CB336701).
		The work of S. Bi was supported in part by the National Natural Science Foundation of China (Project No. 61501303), the Foundation of Shenzhen City (Project No. JCYJ20160307153818306), and the Science and Technology Innovation Commission of Shenzhen (Project No. 827/000212).
		\IEEEcompsocthanksitem
		Yingxiao~Zhang and Ying-Jun Angela Zhang are with the Department of Information Engineering, The Chinese University of Hong Kong, HK. Email:~\{zy012,~yjzhang\}@ie.cuhk.edu.hk.
		\IEEEcompsocthanksitem
		Suzhi~Bi is with the College of Information Engineering, Shenzhen University, Shenzhen,  Guangdong, China. Email:~bsz@szu.edu.cn.
	}
}

\maketitle

\begin{abstract}
In ultra-dense networks (UDNs),  distributed radio access points (RAPs) are configured into small virtual cells around mobile users for fair and high-throughput services.
In this correspondence, we evaluate the performance of user-centric joint transmission (JT) in a UDN with a number of virtual cells.
In contrast to existing cooperation schemes, which assume constant RAP transmit power, we consider a total transmit power constraint for each user, and assume that the total power is optimally allocated to the RAPs in each virtual cell using maximum ratio transmission (MRT).
Based on stochastic geometry models of the RAP and user locations, we resolve the correlation of transmit powers introduced by MRT, and derive the average user throughput.
Numerical results show that user-centric JT with MRT provides high signal-to-noise ratio (SNR) without generating severe interference to other co-channel users.
Moreover, we show that MRT precoding, while requiring channel-state-information (CSI), is essential for the success of JT.
\end{abstract}


\section{Introduction}
\label{sec:introduction}
Driven by the exponential growth of data traffic, wireless mobile networks are experiencing a progressive shift towards a dense deployment of radio access points (RAPs), resulting in ultra-dense networks (UDNs) \cite{2016:Kamel,2015:Bi}.
To avoid severe interference from closely located RAPs, cooperative joint transmission (JT) has been recognized as an important technique, where a set of RAPs are coordinated to transmit the data simultaneously \cite{2017:Qamar}.
With effective joint signal processing, interference within the cooperation cluster is canceled out, and the power of useful signals received at the user is boosted by increased transmission diversity. 
Conventionally, cooperation clusters are pre-assigned by the RAP locations following static patterns, such as three neighbor base stations (BSs) in hexagonal cellular networks \cite{2016:Sucasas}.
Although the intra-cluster interference is successfully mitigated by BS cooperation, the users at cluster edges still suffer from strong inter-cluster interference.
Instead, in user-centric cooperation, a virtual cell is configured on demand by a set of RAPs surrounding a user.
Adaptive cooperation eliminates the cluster edge effect in BS-centric cooperation.
Moreover, joint processing within a few local RAPs in the vicinity of each user saves considerable channel feedback and computational complexity over network-wide joint beamforming.

An important design parameter that affects the performance of user-centric cooperation is the size of virtual cells.
From the perspective of a single user, the performance gain from cooperation increases and gradually saturates as its virtual cell becomes larger \cite{2014:Nigam,2014:Tanbourgi}.
The cooperation gain is even more limited in interference-limited networks, where larger virtual cells create stronger interference to other users.
Reference \cite{2014:Feng} investigates the trade-off between the cooperation gain and the severer interference, and shows that cooperation is preferred for sparse networks.
Reference \cite{2014:Lin} shows that selective transmission from the nearest RAP outperforms blanket JT when channel state information (CSI) is absent at the transmitter.
Reference \cite{2015:Baccelli} studies pairwise BS cooperation and derives a geometric policy to choose cooperation or not.
Reference \cite{2015:Lee} shows that the optimal cluster size is small due to pilot overhead.
Reference \cite{2016:Wang} derives an upper-bound on the average user rate and uses the result to find the optimal virtual cell size.
All the above works assume constant transmit power for each BS.
Without efficient power control, the cooperation gain from large virtual cells is easily overwhelmed by the increased interference.

In this correspondence, we endeavor to find the optimal design of user-centric joint transmission in a UDN with  a number of virtual cells.
In contrast to previous works, which assume constant RAP transmit power, we consider a total transmit power constraint for each user and assume that the power is optimally allocated to the RAPs in each virtual cell using maximum ratio transmission (MRT).
A major challenge brought about by the considered power control is that now the transmit powers of the RAPs in each virtual cell are correlated.
This largely complicates the throughput analysis compared to the case with constant RAP transmit powers.
Based on stochastic geometry models of the RAP and user locations, we resolve the correlation of the transmit powers introduced by MRT precoding, and derive an integral expression of the average user throughput.
From the analysis, we explicitly explain the impact of RAP density and path-loss exponent on the user throughput.
From numerical results, we show that the average user throughput increases with the virtual cell size, but the gain gradually diminishes.
By comparing with other cooperation schemes, we show that MRT, while involving more complexities of CSI feedback and joint precoding, provides high signal-to-noise ratio (SNR) without generating severe interference to other users.
Moreover, when CSI is absent for joint precoding, it is more beneficial, in terms of both throughput and complexity, to select a single transmitter in each virtual cell rather than using multiple RAPs for JT.

\section{System model}
\label{sec:system}
We consider downlink transmission in a UDN with a large number of distributed RAPs.
The locations of RAPs are modeled by a Poisson point process (PPP) $\Phi_\mathrm{r}=\{x_i\}_{i=0}^{\infty}$ with density $\lambda_\mathrm{r}$, where each $x_i\in\mathbb{R}^2$ denotes the location coordinates of RAP $i$.
The PPP model captures the irregular locations of RAPs in the UDN.
For User $k$ located at $u_k\in\mathbb{R}^2$, a set of RAPs located within distance $C$ to the user are configured as its virtual cell,	which is denoted by $\mathcal{C}(u_k)=\{x_i\in \Phi_\mathrm{r}: |x_i-u_k|\le C \}$, where $|\cdot|$ denotes the Euclidean norm.
The cell radius $C$ is a tunable system parameter that controls the span of cooperation.
We assume that all RAPs and users are equipped with a single antenna.

\begin{figure}
	\centering
	\includegraphics[width=0.35\textwidth]{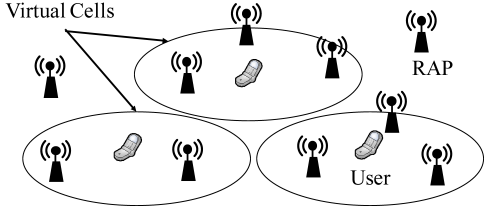}
	\vspace{3pt}
	\caption{Snapshot of three user-centric virtual cells.}
	\label{fig:1}
\end{figure}

We assume that a set of orthogonal channels (e.g., time-slots in TDMA or sub-carriers in OFDM) are allocated to the virtual cells for downlink transmissions.
Specifically, we assume a soft channel reuse pattern, where the same channel is only reused by the virtual cells that are separated by a distance larger than $D$.
In other words, each user is protected by an exclusive region of radius $D$, where no other co-channel users exist.
By setting $D\ge 2C$, the virtual cells of co-channel users are completely separated.
We model the locations of users contending for a channel as a PPP $\Psi_\mathrm{u}=\{u_k\}_{k=0}^\infty$ with intensity $\lambda_\mathrm{u}$.
Let $M$ be the number of users located within distance $D$ to a typical user.
Then, $M$ follows a Poisson distribution with mean $\lambda_\mathrm{u} \pi D^2$.
Since all users are equivalent, the probability that a user is scheduled is given by
\begin{equation}
\begin{aligned}
		p_\mathrm{r}(D) = \sum_{m=1}^\infty \frac{1}{m+1} \mathbf{Pr}\left(M = m\right)
		= \frac{1-e^{-\lambda_\mathrm{u} \pi D^2}}{\lambda_\mathrm{u} \pi D^2}.
		\label{eqn:1}
\end{aligned}
\end{equation}
The selected co-channel users form a hard core point process (HCPP) $\Psi_\mathrm{u}'$, where the minimum separation distance is $D$.
In contrast to the lattice model in \cite{2014:ZhangYX}, which considers the most-packed situation, the HCPP model reflects a random distribution of co-channel users with a guaranteed separation distance.
A snapshot of the network with three virtual cells is presented in Fig. \ref{fig:1}.
For the rest of the paper, we refer to co-channel users as users, unless otherwise specified.

Consider narrow-band flat-fading channels affected by both large-scale path loss and Rayleigh fading.
Let $\ell(d)=\max\left(d, d_0\right)^{-\alpha}$ be the path-loss function, where $d$ is the distance between the transmitter and the receiver, $d_0$ is the reference distance, and $\alpha > 2$ is the path-loss exponent.
This path-loss model takes account of the minimum separation between users and RAPs in real situations \cite{3gpp:COMP}.
Let $g_{ik}\sim \exp(1)$ and $\theta_{ik}\sim \left[0,2\pi\right)$ denote the power gain and phase shift of the Rayleigh fading, respectively, for the link between RAP $i$ and User $k$.
Then, the signal received by User $k$ located at $u_k\in\mathbb{R}^2$ is given by
\begin{equation}
\begin{aligned}
	& Y_k = {\mathsmaller \sum_{x_i\in \mathcal{C}(u_k)}} \sqrt{\ell(|x_i-u_k|) g_{ik} P} e^{j\theta_{ik} } W_{ik} X_k + Z_k 
	\\
	& + {\mathsmaller \sum_{u_{k'}\in \Psi_\mathrm{u}'\setminus \{u_k\}}} {\mathsmaller \sum_{x_i\in \mathcal{C}(u_{k'})}} \sqrt{\ell(|x_i-u_{k}|) g_{ik} P} e^{j \theta_{ik}} W_{ik'}  X_{k'}, 
\end{aligned}
\label{eqn:2}
\end{equation}
where $X_k \sim \mathcal{CN}(0,1)$ is the transmitted signal for User $k$, $P$ is the total transmit power for each user, $W_{ik}\in \mathbb{C}$ is the encoder of RAP $i$ for the transmission to User $k$, and $Z_k\sim \mathcal{CN}(0,\sigma_\mathrm{n}^2)$ is additive white Gaussian noise with power $\sigma_\mathrm{n}^2$.

In each virtual cell, the RAPs form a multiple-input-single-output (MISO) channel to serve the user.
To maximize the received SNR under the total power constraint, MRT is the optimal precoding scheme.
Formally, the MRT encoder for RAP $i$ with $x_i\in \mathcal{C}(u_k)$ is given by $W_{ik} = \sqrt{w_{ik}} e^{-j\theta_{ik}}$, where
\begin{equation}
	w_{ik} = \frac{\ell(|x_i-u_k|) g_{ik}}{\sum_{ x_j\in \mathcal{C}(u_k) } \ell(|x_j-u_k|) g_{jk} }.
	\label{eqn:3}
\end{equation}
For $x_i\notin \mathcal{C}(u_k)$, we have $w_{ik}=0$, indicating that RAP $i$ is not serving User $k$.
In MRT, the total transmit power for each user is optimally allocated to the RAPs according to their instantaneous channel gains.

Without loss of generality, we focus on the performance of a typical user with index $0$ and location at the origin $u_0=o$.
The signal-to-interference-plus-noise ratio (SINR) is given by $\mathrm{SINR} =S/(J+\sigma_\mathrm{n}^2/P)$,	where
\begin{align}
	S &= {\mathsmaller \sum_{x_i\in \mathcal{C}(u_0)}}	\ell(|x_i|) g_{i0},
	\label{eqn:4}
	\\
	J &=  {\mathsmaller \sum_{u_k\in \Psi_\mathrm{u}'\setminus \{u_0\}}} {\mathsmaller \sum_{x_i\in \mathcal{C}(u_k)}} \ell(|x_i|) g_{i0} w_{ik},
	\label{eqn:5}
\end{align}
and $\sigma_\mathrm{n}^2/P$ is the normalized noise power.
We see that the signal power $S$ is a sum of the channel gains for all serving RAPs.
With MRT precoding, the transmission diversity of multiple RAPs in each virtual cell is transformed into a power boost of desired signals at the receiver.
In contrast, the interference power $J$ depends not only on the interfering channels, but also on the transmit power weights of the RAPs, i.e., $w_{ik}$.
Note that the power weights of the RAPs in each virtual cell sum up to $1$ due to power normalization in MRT precoding.

We evaluate the performance of user-centric JT by the average throughput of a typical user scheduled in a channel with bandwidth $B$, which is defined as
\begin{equation}
	\tau = \mathbf{E}\left[B\log_2\left(1+\mathrm{SINR}\right)\right].
	\label{eqn:6}
\end{equation}
In (\ref{eqn:6}), the expectation is taken with respect to all fast fadings and user and RAP locations.

From a system perspective, we are also interested in the average spatial throughput, which is defined as
\begin{equation}
	\eta = \lambda_\mathrm{u} p_\mathrm{r}(D)  \tau(D),
	\label{eqn:7}
\end{equation}
where $ \lambda_\mathrm{u} p_\mathrm{r}(D) $ is the average number of co-channel users per km${}^2$.
Intuitively, as $D$ increases, each scheduled user would obtain higher throughput, while fewer users can be scheduled in the same channel.
Hence, it is important to evaluate $\eta$ as a function of $D$ for the overall system throughput.

\section{Throughput Analysis}
\label{sec:analysis}
In this section, we derive the average user throughput $\tau$, which depends on the distributions of the signal power $S$ and interference power $J$.
Due to the geographical separation of the virtual cells,  $S$ and $J$ are independent.
Hence, we start by characterizing $S$ and $J$ separately, and then use the results to derive $\tau$.

\subsection{Signal Power}
\label{sec:signal}
From the definition in (\ref{eqn:4}), we see that $S$ is a sum over a subset of the RAPs in $\Phi_\mathrm{r}$.
Treating the Rayleigh fading $g_{i0}$ as an independent mark of each RAP, we can characterize the distribution of $S$ by the Laplace transform \cite{2017:ElSawy}
\begin{equation}
\begin{aligned}
	&	\mathcal{L}_S(t) = \mathbf{E}_{\Phi_\mathrm{r}, g}\left[e^{-t \sum_{x_i\in \mathcal{C}(u_0)}	\ell(|x_i|) g_{i0} } \right]
	\\
	&\quad = \exp\left(-2\pi\lambda_\mathrm{r} \int_{r=0}^C 
	\left\{1- \mathbf{E}_{g_{i0}} \left[e^{-t \ell(r) g_{i0} } \right] \right\} rdr  \right).
\end{aligned}
\label{eqn:8}
\end{equation}
With the exponential distribution of $g_{i0}$, we have
\begin{equation}
\mathcal{L}_S(t) = \exp\left[-2\pi\lambda_\mathrm{r} \int_{r=0}^C \frac{t\ell(r)}{1+t\ell(r)}  rdr \right].
\label{eqn:9}
\end{equation}
From (\ref{eqn:9}), we observe that $S$ increases as the size of virtual cells $C$ becomes larger, even though the total transmit power is unchanged in the virtual cell.
Note that while (\ref{eqn:9}) gives an explicit expression of $\mathcal{L}_S(t)$, the probability density function (PDF) and cumulative distribution function (CDF) of $S$ are not in closed-form for a general path-loss function.
Hence, we use $\mathcal{L}_S(t)$ directly to calculate $\tau$ in Section \ref{sec:rate}.

\subsection{Power Weights in MRT}
\label{sec:weight}
In this subsection, we analyze the transmit powers in MRT, which are needed in the analysis of interference.
From (\ref{eqn:5}), we see that interference from a RAP is determined not only by the channel fading to a typical user, but also by the RAP's transmit power.
With MRT precoding, the transmit powers of the cooperating RAPs are correlated and sum up to $P$, which makes the analysis of $w_{ik}$ very difficult.
For analytical tractability, we use the conditional first-order moment of $w_{ik}$, which represents the average transmit power for a given RAP.

\begin{lemma}
	\label{lmm:1}
	For a RAP located at distance $r\le C$ to a user, the first-order moment of its power weight, i.e., 
	$\mathbf{E}\left[w_{ik} \big| |x_i-u_k|=r\right]$, is given by
	\begin{equation}
		\mathcal{W}(r) =\int_0^\infty \frac{1}{(1+t)^2} \mathcal{L}_{S}\left[\frac{t}{\ell(r)}\right]  \,dt,
		\label{eqn:10}
	\end{equation}
	where $\mathcal{L}_S(t)$ is given by (\ref{eqn:9}).
\end{lemma}
\begin{IEEEproof}
See Appendix A.
\end{IEEEproof}

Lemma \ref{lmm:1} characterizes the power weight for a RAP as a function of its distance to the served user.
From (\ref{eqn:10}), we see that $\mathcal{W}(r)$ is a decreasing function of $r$,
	indicating that a RAP located farther away from the user is allocated less transmit power.
Moreover, as the cell size $C$ increases, $\mathcal{L}_S(t)$ decreases and thus $\mathcal{W}(r)$ becomes smaller.
This is because under the total power constraint, a RAP is allocated less transmit power when more RAPs are included in the virtual cell.
In the following calculation of the interference power, we use a first-order approximation,
	where $w_{ik}$ is treated as a constant weight with value $\mathcal{W}(|x_i-u_k|)$.

\subsection{Interference Power}
\label{sec:intf}
Now we calculate the overall interference $J$ as defined in (\ref{eqn:5}). 
Let $I_k=\sum_{x_i\in \mathcal{C}(u_k)} \ell(|x_i|) g_{i0} w_{ik}$ be the interference power from the virtual cell of User $k$.
Then (\ref{eqn:5}) can be written as $J=\sum_{k\neq 0} I_k$.
Due to the geographical separation of virtual cells, the $I_k$'s are independent.

We see that $J$ is a sum over the users in the HCPP $\Psi_\mathrm{u}'$ that are separated by a minimum distance $D$.
The HCPP is difficult to analyze since the locations of points are no longer independent.
It is stated in \cite{2017:ElSawy} that points located at distances larger than $D$ can be approximately modeled as a PPP when calculating their aggregated interference.
By the PPP approximation, we derive the Laplace transform of $J$ as
\begin{equation}
	\begin{aligned}
	&\mathcal{L}_J(t) = \mathbf{E} \left( e^{-t \sum_{u_k\in \Psi_\mathrm{u}'\setminus \{u_0\}} I_k}\right)
	\\
	&\quad \approx \exp\left\{-2\pi\lambda_\mathrm{u} p_\mathrm{r}(D) 
	\int_{\rho=D}^{\infty} \left[1- \mathcal{L}_{I(\rho)}(t) \right]\rho d\rho \right\},
	\end{aligned}
	\label{eqn:11}
\end{equation}
where $ \mathcal{L}_{I(\rho)}(t) =  \mathbf{E} \left(e^{-t I_k} \big| |u_k| = \rho\right)$.
By the first-order approximation of $w_{ik}$, we have
\begin{equation}
\begin{aligned}
	&\mathcal{L}_{I(\rho)}(t) \approx \mathbf{E} \left[e^{-t \sum_{x_i\in \mathcal{C}(u_k)} \ell(|x_i|) g_{i0} \mathcal{W}(|x_i-u_k|) }\Big ||u_k|=\rho\right]
	\\
	&=\exp\left[-\lambda_\mathrm{r} 	\int_{0}^{2\pi} \int_{0}^C
	\frac{t\ell'(\rho, r, \theta)\mathcal{W}(r)}
	{1+t\ell'(\rho, r, \theta)\mathcal{W}(r)}  rdrd \theta \right],
\end{aligned}
\label{eqn:12}
\end{equation}
	where $\ell'(\rho, r, \theta)= \ell\left(\sqrt{\rho^2+r^2-2r\rho\cos\theta}\right)$.
Substituting (\ref{eqn:12}) into (\ref{eqn:11}) gives the final expression of $\mathcal{L}_J(t)$.

\subsection{Far-Field Approximation}
When $|u_k|\gg C$, $\ell(|x_i|) \approx \ell(|u_k|)$ for any $x_i \in \mathcal{C}(u_k)$, meaning that the RAPs in a far-away virtual cell share similar path losses.
Hence, the aggregated interference can be treated as that from a single transmitter, i.e., $I_k \approx \ell(|u_k|) \tilde{g}_k$, where $\tilde{g}_k=\sum_{x_i\in \mathcal{C}(u_k)} g_{i0}$ is the power gain of the small-scale fading.
By treating $\tilde{g}_k$ as a random Rayleigh fading component, we can simplify (\ref{eqn:12}) to
\begin{equation}
\begin{aligned}
	\mathcal{L}_{I(\rho)}(t) &\approx \mathbf{Pr}\left[\mathcal{C}(u_k)=\emptyset \right]
	+ \mathbf{E}\left[ e^{-t \ell(\rho) \tilde{g}_k} \right]\mathbf{Pr}\left[\mathcal{C}(u_k)\neq \emptyset \right]
	\\
	&\approx  e^{-\lambda_\mathrm{r} \pi C^2} + \left(1-e^{-\lambda_\mathrm{r} \pi C^2} \right) \frac{1}{1+t\rho^{-\alpha}}.
\end{aligned}
\label{eqn:13}
\end{equation}
Note that this simple far-field approximation comes at a cost of accuracy, where the diversity due to multiple interfering RAPs in a virtual cell is underestimated.

To keep a balance between accuracy and complexity in calculating the Laplace transform of $J$, we divide the overall interference field into near and far fields by some distance $d\gg D$.
Let $J^\mathrm{n}$ be the aggregated interference power from the near field, i.e., those virtual cells of users located within $d$, and $J^\mathrm{f}$ be that from the far field, i.e., those virtual cells of users located farther away than $d$.
Then, we have $\mathcal{L}_J(t) = \mathcal{L}_{J^\mathrm{n}}(t) \mathcal{L}_{J^\mathrm{f}}(t) $. 
On one hand, the near-field interference can be computed accurately with the expression of $ \mathcal{L}_{I(\rho)}(t)  $ (\ref{eqn:12}).
On the other hand, the aggregated interference from a large number of virtual cells in the far field can be computed efficiently with the far-filed approximation (\ref{eqn:13}), which is given by
\begin{equation}
\begin{aligned}
	&\mathcal{L}_{J^\mathrm{f}}(t) = \exp\left\{-2\pi\lambda_\mathrm{u} p_\mathrm{r}(D)\int_{\rho=d}^{\infty} \left[1- \mathcal{L}_{I(\rho)}(t) \right]\rho d\rho \right\}
	\\
	&\approx \exp\left[-2\pi\lambda_\mathrm{u} p_\mathrm{r}(D) \left(1- e^{-\lambda_\mathrm{r}\pi C^2 }\right) \int_{\rho=d}^{\infty} \frac{t\rho}{\rho^{\alpha}+t} d\rho \right].
\end{aligned}
\label{eqn:14}
\end{equation}
We discuss the accuracy of the approximation with numerical results in Section \ref{sec:simulation}.

\subsection{Average User Throughput}
\label{sec:rate}
In Section \ref{sec:signal} and \ref{sec:intf}, we have characterized the distributions of $S$ and $J$ by their Laplace transforms, respectively.
In general, distribution functions such as PDF and CDF, are not given in closed form, making it difficult to derive the distribution of SINR.
Instead, we derive the average user throughput $\tau$ directly from the Laplace transforms.

\begin{theorem}
	\label{thm:1}
	The average throughput of a typical user as defined in (\ref{eqn:6}) is given by
	\begin{equation}
		\tau = \frac{B}{\ln 2}\int_{t=0}^\infty e^{-t\sigma_\mathrm{n}^2/P} \mathcal{L}_{J}(t) \left[1-\mathcal{L}_{S}(t)\right] \frac{1}{t} dt,
		\label{eqn:15}
	\end{equation}
	where $\mathcal{L}_{S}(t)$ and $\mathcal{L}_{J}(t)$ are given in (\ref{eqn:9}) and (\ref{eqn:11}), respectively.
\end{theorem}
\begin{IEEEproof}
	See Appendix B.
\end{IEEEproof}

Note that Theorem \ref{thm:1} provides a way to compute the average user throughput achieved by MRT precoding in an UDN with virtual cells.
The system parameters are implicitly included in the Laplace transforms.
When $S$ and $J$ are bounded, the integration in (\ref{eqn:15}) can be numerically evaluated.

\emph{Impact of path-loss exponent:}
With a small path-loss exponent $\alpha$, the transmitted signals undergo milder path loss, resulting in both high signal power and high interference power.
As $\alpha\rightarrow 2$, the integration in the exponential term of (\ref{eqn:14}) goes to infinity, and thus $\mathcal{L}_{J^\mathrm{f}}(t)=0$.
Together with $\mathcal{L}_S(t)< 1$, we conclude that $\lim_{\alpha\rightarrow 2} \tau= 0$.
This is because the aggregated interference from infinite area is unbounded in a mild fading environment.

\emph{Impact of RAP density:}
As $\lambda_\mathrm{r}$ increases, more RAPs are involved for JT in each virtual cell,
	resulting in more transmission diversity for both the signal and interference.
As $\lambda_\mathrm{r}\rightarrow \infty$, $\mathcal{L}_S(t)\rightarrow 0$, while $\mathcal{L}_J(t)> 0$.
As a result, $\tau\rightarrow \infty$, meaning that the average user throughput can be improved by network densification.

The impact of other system parameters, such as $C$ and $D$, cannot be directly observed from (\ref{eqn:15}).
In the next section, we evaluate $\tau$ numerically and obtain more insight on the design of virtual cells.

\section{Numerical Results and Discussion}
\label{sec:simulation}
In this section, we present numerical evaluations of the analytical results derived in Section \ref{sec:analysis}.
In the simulations, we drop the RAPs and users in a square area with side lengths of $10$ km.
The density of RAPs is $\lambda_\mathrm{r}=50$ km${}^{-2}$, and the density of users is $\lambda_\mathrm{u}=20$ km${}^{-2}$.
The minimum separation of co-channel users is set to $D=0.4$ km,
	and the virtual cells are formed around co-channel users with radius $C=0.2$ km, unless otherwise specified.
A standard path-loss model with $\alpha=3.6$ and $d_0=10$ m is adopted.
The channel bandwidth $B=10$ MHz,
	and the noise power spectral density is $-174$ dBm \cite{3gpp:COMP}.
The total transmit power for each user is $P=24$ dBm.
All simulation results are obtained by averaging over 1000 independent locations of users and RAPs, each with 100 Rayleigh fading realizations.

\begin{figure}
	\centering
	\includegraphics[width=0.40\textwidth]{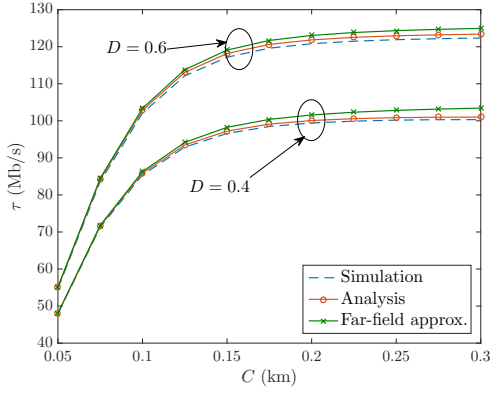}
	\caption{Average user throughput $\tau$ versus virtual cell size {\small $C$ \normalsize}.}
	\label{fig:2}
\end{figure}

\begin{figure}
	\centering
	\includegraphics[width=0.40\textwidth]{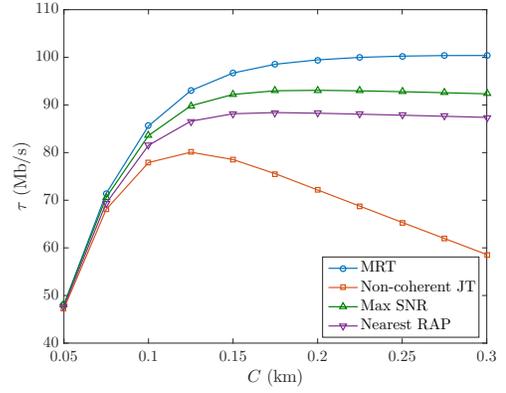}
	\caption{Comparison of the average user throughput using MRT with that of using other cooperation schemes.}
	\label{fig:3}
\end{figure}

\subsection{Validation of Analysis}
In Fig. \ref{fig:2}, we plot the average user throughput $\tau$ as a function of the cell radius $C$, and compare the analytical results calculated by Theorem \ref{thm:1} and by the far-field approximation in (\ref{eqn:14}) with empirical results.
We extend our analytical results to the case $C>D/2$ where some virtual cells may overlap.
We can see that the analytical results match the empirical results fairly well for various settings.
Moreover, the far-field approximation provides a tight upper bound, where the approximation error increases as the virtual cells become larger.
This indicates that the interference is underestimated if a virtual cell is treated as a single interferer at the cell center.

\subsection{Impact of virtual cell size}
In Fig. \ref{fig:2}, we can see that $\tau$ increases with $C$, but the gain gradually saturates.
This is different from \cite{2014:Lin,2014:Feng}, which show that small virtual cells are preferred when each point transmits with a constant power.
This is because MRT is able to provide high cooperation gain without generating severe interference to other co-channel users.
In practice, the operational complexity of MRT, including channel estimation, CSI exchange, synchronization, and joint precoding, increases with the number of cooperating RAPs.
Therefore, the design of virtual cell size should achieve a balance between performance and complexity.

\subsection{Impact of cooperation scheme}
In Fig. \ref{fig:3}, we compare MRT with other cooperation schemes, i.e., non-coherent JT, max SNR, and nearest RAP, by their achieved average user throughputs.
In contrast to MRT, non-coherent JT requires no CSI, where the cooperating RAPs transmit the same data without any phase adjustment or power control \cite{2014:Tanbourgi}.
We also simulate selective transmission schemes as benchmarks, where a single RAP in each virtual cell is selected as the transmitter while other RAPs cooperate by keeping silent.
Specifically, when CSI is available at the transmitter, the RAP providing maximum SNR is selected; otherwise, the nearest RAP is assigned as the transmitter in each virtual cell.
Compared to JT with multiple transmitting RAPs, selective transmission has much lower complexity.
From Fig. \ref{fig:3}, we see that MRT outperforms all other cooperation schemes.
An interesting observation is that non-coherent JT has even lower throughput than selective transmission with the nearest RAP.
This implies that joint precoding and power control are necessary for the success of multiple-RAP JT.
Otherwise, selective transmission should be adopted.

\subsection{Average spatial throughput}
\begin{figure}
	\centering
	\includegraphics[width=0.40\textwidth]{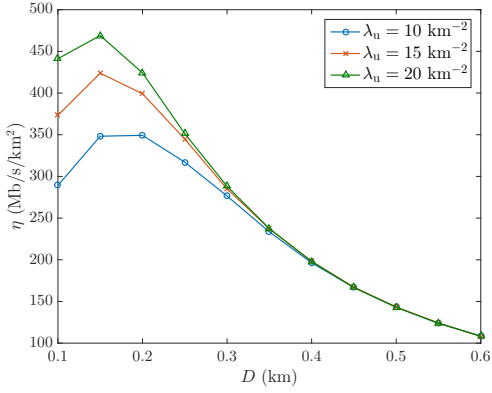}
	\caption{Average spatial throughput $\eta$ versus minimum separation distance of co-channel users 
		{\small $D$ \normalsize}.}
	\label{fig:4}
\end{figure}
From Fig. \ref{fig:2}, we see that the average user throughput $\tau$ increases with $D$.
In other words, each user obtains a higher throughput if the co-channel users are farther separated.
However, from a system perspective, fewer users are served in the same channel.
In Fig. \ref{fig:4}, we plot the average spatial throughput $\eta$, which is defined in (\ref{eqn:7}), as a function of $D$ for fixed user contention intensities $\lambda_\mathrm{u}$.
Specifically, the cell size is set to $C=D/2$ for a given $D$.
From Fig. \ref{fig:4}, we see that $\eta$ firstly increases and then decreases quickly with $D$.
The optimal $D$ that maximizes $\eta$ depends on the contention intensity $\lambda_\mathrm{u}$.
When there is a large number of users contending for the channel, it is better to choose a smaller $D$, such that more users are admitted while the throughput of each user is limited.

\section{Conclusions}
\label{sec:conclusion}
In this correspondence, we have examined the performance of user-centric JT with virtual cells in an UDN.
Specifically, we assumed that MRT is applied in each virtual cell to maximize the SNR received by each user.
From our analysis, we have shown that user-centric MRT can provide high cooperation gain to each user without causing severe interference to other co-channel users.
Moreover, we have shown that channel-dependent joint precoding, while increasing the complexity of coordination, is essential for RAP cooperation.
Otherwise, the interference from cooperating RAPs would easily overwhelm the cooperation gain, resulting in severe loss of system throughput.
In turn, the centralized architecture with seamless coordination is a key technical enabler for the success of multiple RAP JT.

\appendices
\section*{APPENDIX A: Proof of Lemma \ref{lmm:1}}
\label{appx:lmm:1}
Consider the virtual cell of User $k$.
Given that RAP $i$ is located at $x_i$ such that $|x_i-u_k|=r$, its power weight $w_{ik}$ can be calculated by definition (\ref{eqn:3}).
Let $S'=\sum_{x_j \in \mathcal{C}(u_k)\setminus\{x_i\}} \ell(|x_j-u_k|) g_{jk}$ denote the received signal power from the RAPs other than RAP $i$.
Then, we can compute the first-order moment as
\begin{equation}
\begin{aligned}
	& \mathcal{W}(r)=\mathbf{E}_{g_{ik}, S'} \left[\frac{\ell(r) g_{ik}}{\ell(r) g_{ik} + S'} \right] 
	\\
	&\quad \stackrel{\mathrm{(a)}}{=} \mathbf{E}_{g_{ik}} \left\{ \ell(r) g_{ik} \int_0^\infty \mathbf{E}_{S'}\left[e^{-t [\ell(r) g_{ik} + S']}\right] dt \right\}
	\\
	&\quad \stackrel{\mathrm{(b)}}{=} \int_0^\infty \mathbf{E}_{g_{ik}}\left(g_{ik}  e^{-t g_{ik} }\right)\mathbf{E}_{S'} \left[e^{-tS'/\ell(r)}\right] dt,
\end{aligned}
\label{eqn:16}
\end{equation}
where (a) follows from the the independence of $S'$ and $g_{ik}$ and the mathematical result $\mathbf{E}(Z^{-n})= \Gamma(n)^{-1} \int_{t=0}^\infty t^{(n-1)} \mathbf{E}(e^{-t Z}) dt$ \cite{1981:Cressie}, and (b) follows from the Tonelli's theorem.
By the exponential distribution of $g_{ik}$, we have $\mathbf{E}_{g_{ik}}\left(g_{ik}  e^{-t g_{ik} } \right)= \left(1+t \right)^{-2} $.
By the Slivnyak-Mecke Theorem \cite{2017:ElSawy}, we have $\mathbf{E}_{S'} [e^{-t S'/\ell(r)}] = \mathcal{L}_S\left[t/\ell(r)\right]$, which gives the final expression in (\ref{eqn:10}).

\section*{APPENDIX B: Proof of Theorem \ref{thm:1}}
\label{appx:thm:1}
By definition (\ref{eqn:6}), we have
\begin{equation}
\begin{aligned}
	\tau &= \frac{B}{\ln 2} \mathbf{E}_{S,J}\left[\ln\left(1+\frac{S}{J+\sigma_\mathrm{n}^2/P}\right) \right]
	\\
	&\stackrel{\mathrm{(a)}}{=} \frac{B}{\ln 2} \mathbf{E}_{S,J}\left[\int_{t=0}^\infty e^{-t} \left(1-e^{-t \frac{S}{J+\sigma_\mathrm{n}^2/P}} \right) \frac{1}{t} dt\right]
	\\
	&=\frac{B}{\ln 2} \mathbf{E}_{S,J} \left[\int_{t=0}^\infty e^{-t (J+\sigma_\mathrm{n}/P)} \left(1-e^{-tS} \right) \frac{1}{t} dt\right]
	\\
	&\stackrel{\mathrm{(b)}}{=} \frac{B}{\ln 2}\int_{t=0}^\infty e^{-t \sigma_\mathrm{n}/P} \mathbf{E}(e^{-tJ}) \left[1-\mathbf{E}(e^{-tS}) \right] \frac{1}{t} dt,
\end{aligned}
\label{eqn:17}
\end{equation}
where (a) follows from \cite[Lemma 1]{2008:Hamdi}, and (b) is obtained by the independence of $S$ and $J$.
With the Laplace transform expressions, we then obtain the expression in (\ref{eqn:15}).



\end{document}